\documentclass[aps,prl,twocolumn,showpacs,superscriptaddress]{revtex4}
\usepackage{graphicx}
\usepackage{hyperref}
\usepackage{natbib}
\usepackage{bm}
\begin{document}

\title{Destruction of the Kondo effect by a local measurement}
\author{M. I. Katsnelson}
\affiliation{Ames Laboratory, Iowa State University, Ames IA 50011, USA}
\affiliation{Department of Physics, Uppsala University, Box 530,
SE-75121, Uppsala, Sweden}
\author{V. V. Dobrovitski}
\affiliation{Ames Laboratory, Iowa State University, Ames IA 50011, USA}
\author{H. A. De Raedt}
\affiliation{Institute for Theoretical Physics and Materials 
  Science Centre, University of Groningen, Nijenborgh 4, NL-9747 AG
  Groningen, The Netherlands}
\author{B. N. Harmon}
\affiliation{Ames Laboratory, Iowa State University, Ames IA 50011, USA}

\begin{abstract}
We show that the local spin measurement which decoheres
the localized spin in a Kondo system, suppresses the
Abrikosov-Suhl resonance and destroys the Kondo effect. This
happens due to elimination of the entanglement between the localized
spin and the conduction electrons, and differs essentially from
smearing of the resonance by dissipation. Considering decoherence
by a spin bath, we predict that the Kondo effect
disappears when the Kondo temperature becomes smaller than the
coupling with a bath. This effect can be detected in
experiments on ``quantum corrals'' or quantum dots doped by
impurities with internal degrees of freedom.
\end{abstract}

\date{\today}
\pacs{03.65.Yz; 75.20.Hr; 73.21.La}

\maketitle

The Kondo effect discovered originally as a mechanism to explain
the resistivity minimum in dilute magnetic alloys \cite{kondo} is
one of most interesting many-body phenomena in condensed matter
physics. It plays a crucial role for heavy fermion
physics \cite{hewson}, for metallic glasses and systems with
orbitally degenerate ions \cite{cox}, for quantum
dots \cite{qdtheory,qdexp}, for quantum tunneling in metals
\cite{kagan}, etc. The key feature of the Kondo
effect is the formation of the Abrikosov-Suhl resonance near the
Fermi level. The STM technique allowed visualization of
this resonance for
magnetic impurities \cite{STM,coral} and for orbital 
Kondo effect on atomically clean metal surfaces \cite{kol}.
This resonance is a consequence of the strong
quantum correlations between the localized magnetic moment (or
other center with internal degrees of freedom) and the
conduction electrons. These two subsystems form a singlet
state analogous to the state of the Einstein-Podolsky-Rosen spin
pair \cite{book,zurek,joos}. Such states are often fragile, since the
quantum correlations can be destroyed by decoherence,
as a result of quantum measurement of one of the
spins (or both) either by a specially designed device or by
an environment which has the same effect as a measurement apparatus.

In this paper we show that a spin measurement, which decoheres
the Kondo system and destroys correlation between the localized
spin and the conduction electrons, suppresses the Abrikosov-Suhl
resonance and thus leads to disappearance of the Kondo effect.
This effect is essentially different from the suppression of the
Kondo effect by dissipation (resulting e.g.\ from external
radiation) \cite{microwave}; the crucial difference between
decoherence and dissipation has been discussed in detail
\cite{book,zurek,joos}. We show that for Kondo systems this
leads to definite predictions, which can be tested in
experiments.

One of the standard considerations of the Kondo effect is
based on the $s$--$d$ exchange Hamiltonian \cite{vons}:
\begin{equation}
H_{sd}=\sum_{k,\alpha} \epsilon_{k} c^\dagger_{k\alpha}c_{k\alpha}
 + H_1,\quad H_1=J_{sd} {\bf S}_1 {\bf S}_2
\label{kasham}
\end{equation}
where ${\bf S}_1$ is the impurity spin and 
${\bf S}_2=(1/2N)\sum_{kk'} c^\dagger_{k\alpha}{\bm{\sigma}}_{\alpha\beta}
c_{k'\beta}$ is the spin of the conduction electrons at the
impurity location, $N$ is the number of sites. However,
it is instructive to start first from another, exactly solvable, 
model of the Kondo effect.
Instead of the central spin 1/2, let us consider a central
orbitally degenerate ion, whose internal electrons
(``f''-electrons) are coupled with the conduction electrons,
i.e.\ the infinite-U degenerate Anderson model with the Hamiltonian 
\cite{hewson,GS}
\begin{eqnarray}
H&=&P \sum_{k,\nu} \left[ \epsilon_k c_{k\nu}^{\dag}c_{k\nu}
  +\epsilon_f d_{\nu}^{\dag}d_{\nu}\right] P  + H_1 \\
  \nonumber
H_1&=&P \sum_{k,\nu} \left[V (c_{k\nu}^{\dag}d_{\nu} + d_{\nu}^{\dag}c_{k\nu})\right] P
\label{hamil}
\end{eqnarray}
where $c_{k\nu}$, $d_{\nu}$ are the Fermi operators for 
conduction and localized $f$-electrons, correspondingly, 
$\epsilon_{k}$ and $\epsilon_{f}$ 
are their bare energies counted from the Fermi level, $V$ is
the hybridization parameter, $\nu =1,2,...,N_{f}$ is the 
``flavor'' index
and $P$ is the projection operator into the space with 
$n_{f}=\sum_{\nu}d_{\nu }^{\dag}d_{\nu }<2$.
Both Hamiltonians (\ref{kasham}) and (\ref{hamil})
describe essentially the same physics, so that our conclusions
remain qualitatively valid for both cases, but the latter one
is exactly solvable in the
limit $N_{f}\to\infty$, assuming that $V\to 0$ 
and $V^2N_f=const$. Its ground state wave function is
\cite{GS}
\begin{eqnarray}
\left| \Psi _{0}\right\rangle &=&A\left( \left| 0\right\rangle 
  +|1\rangle \right)  \nonumber \\
\left| 0\right\rangle &=&\prod\limits_{\nu ;k<k_{F}}
 c_{k\nu }^{\dag}\left|
vac\right\rangle  \nonumber \\
\left| 1\right\rangle &=&\sum\limits_{k<k_{F}}|k\rangle \equiv 
\frac{1}{\sqrt{N_{f}}} \sum_{\nu,k<k_{F}} a_k d_{\nu}^{\dag}
 c_{k\nu}|0\rangle  
\label{gs}
\end{eqnarray}
where $|vac\rangle $ is the vacuum state, 
$A=\sqrt{1-\langle n_f\rangle}$ is the normalization factor, 
$\left\langle n_{f}\right\rangle $ is the average occupation of the 
$f$-level, $a_{k}=V\sqrt{N_{f}}%
/\left( E_{0}+\epsilon _{k}-\epsilon _{f}\right)$, and
$E_0$ is the ground state
energy counted from the energy of the ``Fermi sea'' state 
$\left|0\right\rangle$: 
\begin{equation}
E_{0}=\Gamma \left( E_{0}\right) \equiv V^{2}N_{f}\sum\limits_{k<k_{F}}\frac{%
1}{E_{0}+\epsilon _{k}-\epsilon _{f}}.
\label{energy}
\end{equation}
The Green's function of the $f$ electrons $G(E)
=\left\langle \left\langle d_{\nu }|d_{\nu }^{+}\right\rangle
\right\rangle _{E}$ has a pole at the energy $E^{\ast }=\epsilon _{f}-E_{0}$
with the residue $Z=1-\left\langle n_{f}\right\rangle $ which corresponds to
the Abrikosov-Suhl resonance; the energy $E^{\ast }=\rho V^{2}N_{f}\left(
1-\left\langle n_{f}\right\rangle \right) /\left\langle n_{f}\right\rangle $
($\rho $ is the density of states for conduction electrons) plays the role
of the Kondo temperature provided that $Z\ll 1;$ in that regime $Z$ is
exponentially small in $V^{2}N_{f}$ (for more details, see, e.g., \cite
{GS,hewson}). This Abrikosov-Suhl resonance results in the Fano
(anti)resonance in the conduction electron spectrum because of the
identity 
\begin{equation}
\langle\langle c_{k\nu }|c_{k'\nu }^{\dag}\rangle\rangle_E
  = \frac{1}{E-\epsilon_k}+\frac{G\left( E\right) }{\left(
E-\epsilon _{k}\right) \left( E-\epsilon_{k'}\right) }
\label{Fano}
\end{equation}
It is the Fano antiresonance that is observed in the STM experiments
\cite{STM,coral}.

The state (\ref{gs}) describes quantum correlations between the
$f$-electrons and the conduction electrons. These two
subsystems can be considered as an ``EPR pair'' \cite{book},
where the state of the conduction electrons is determined by the
state of the $f$-electrons (and vice versa). One of the most
impressive features of such states is that the decohering action
applied to one subsystem immediately affects the state of the
other subsystem, which has not been directly subjected to
decoherence. We demonstrate below that in our case, such an
influence, e.g.\ the measurement of the number of the 
$f$-electrons $n_f$, leads to an immediate change of the state of the
conduction electrons, and suppresses the Kondo effect.
This effect is similar to the situations explored recently
for Anderson localization \cite{gurvitz}, in the
Bose-Einstein condensate \cite{KDHA},
and for an antiferromagnet \cite{KDHB}. However, the case of 
Kondo systems may be more easily studied in real experiments
(see the discussion below).

Let us assume that the decohering influence of the apparatus
is so effective that it can be described as a von Neumann's
measurement \cite{neumann}. This means that the
initial density matrix $\rho _{i}=\left| \Psi _{0}\right\rangle \left\langle
\Psi_{0}\right|$ is momentarily transformed
into the final one $\rho _{f}=A^{2}\left( \left|
0\right\rangle \left\langle 0\right| +\left| 1\right\rangle \left\langle
1\right| \right)$. The Green's function of the $f$-electrons
after the measurement is
\begin{equation}
G_{f}(E) = -i \int_{0}^\infty dt e^{itE} \mathop{\rm Tr} 
  \rho_f \left[ d_{\nu}(t)d_{\nu}^{\dag} + 
  d_{\nu}^{\dag}d_{\nu}(t) \right],   
\label{green}
\end{equation}
and can be evaluated similarly to Ref.\ \cite{GS}.
One has to introduce the functions $%
\exp (-iHt)\left| \phi \right\rangle $ where $\left| \phi \right\rangle
=\left| 0\right\rangle ,\left| k\right\rangle ,d_{\nu }^{+}\left|
0\right\rangle ,$ write the equations of motion for them with taking into
account only the leading terms in $1/N_{f}$ and make the Laplace
transformation.
% , obtaining:
% \begin{eqnarray}
% \nonumber
% \frac{1}{E-H} |0\rangle &=& \frac{1}{E-\Gamma(E)}
%   \left(|0\rangle + \sum_{k<k_F}\frac{V\sqrt{N_f}}
%   {E+\epsilon_{kf}} |k\rangle\right)\\ 
%   \nonumber
% \frac{1}{E-H}|k\rangle &=& \frac{1}{E+\epsilon_{kf}}
%   \left(|k\rangle + V\sqrt{N_f}\frac{1}{E-H}|0\rangle \right)\\
% \frac{1}{E-H}d_{\nu }^{+}\left| 0\right\rangle &=&
%   \frac{1}{E-\epsilon_f}d_{\nu}^{\dag} |0\rangle,
% \label{resolvent}
% \end{eqnarray}
% where $\epsilon_{kf}=\epsilon_k-\epsilon_f$.
As a result, the Green's function after
the measurement becomes (in the limit of large $N_{f}$): 
\begin{eqnarray}
G_{f}\left( E\right) &=& \frac{A^{2}}{E-\epsilon _{f}+\Gamma \left( \epsilon
_{f}-E\right) }\\ \nonumber
&=& G\left( E\right) \left[ 1+\sum\limits_{k<k_{F}}\frac{%
V^{2}N_{f}}{\left( E_{0}+\epsilon _{k}-\epsilon _{f}\right) 
\left( \epsilon_{k}-E\right) }\right] ^{-1}  
\label{final}
\end{eqnarray}
It has the same pole $E^{\ast }$ as the Green's function before the
measurement, but the residue equals to $\left( 1-\left\langle
n_{f}\right\rangle \right) ^{2\text{ }}$ instead of $1-\left\langle
n_{f}\right\rangle$. It means that the amplitude of the Abrikosov-Suhl
resonance, and, consequently, the Fano antiresonance in the conduction
electron spectra, diminish after the measurement 
by the factor of order of $E^{\ast }/\rho
V^{2}N_{f}$, which is very small in the Kondo regime. 

Experimentally, this effect can be checked with the ``quantum
corral'' setup \cite{coral}, by putting, e.g., a cerium atom on the
metallic surface in the focus of an elliptic ``quantum coral''.
Due to interaction with the $f$-electrons of Ce, the spectrum of
conduction electrons exhibits 
the Fano (anti)resonance
which can be observed by an STM tip placed at
the other focus of the elliptic corral. The charge state of the
Ce atom can be measured, e.g.\ by a point
contact, placed near the atom (as has been analyzed 
in Ref. \cite{zeno}). 
Immediately after the measurement of the Ce atom,
the amplitude of the
Fano resonance should drop drastically.
% the propagation of the decohering action of the measurement from
% one focus of the corral to the other is a particular case of a 
% ``decoherence
% wave'' \cite{KDHA}. 
Instead of
the Ce atom, a magnetic impurity can also be used,
but it seems easier to measure the charge of the ion rather
than its magnetic moment.
Another possibility is to employ the optically induced Kondo
effect, which could be generated by a circularly polarized 
light in a system with an impurity level located in the Fermi sea,
e.g. in a Si-doped GaAs/AlGaAs superlattice with Be impurity,
as has been analyzed in Ref.\ \cite{shah}. The state of the
impurity spin can be measured by a circularly polarized
probe beam: once the probe photon is
absorbed, the state of the impurity is determined uniquely by
the angular momentum conservation. The disappearance
of the Fano resonance can be detected, e.g.\ by X-ray
absorption.

But it might be more feasible to employ the Kondo
effect in quantum dots \cite{qdtheory,qdexp}, where the bath
of environmental degrees of freedom decoheres the central spin, thus
working in essentially the same manner as a measuring device.
Indeed, as the results above show, the spectral weight of the
Abrikosov-Suhl resonance is determined by the non-zero
value of the nondiagonal element
of the density matrix 
$\langle 0|\rho |1\rangle=\langle 0|\Psi_0\rangle\langle 
\Psi_0|1\rangle\neq 0$ (see Eq.\ \ref{gs}).
The interaction ${\cal V}$ between 
the dot and the bath reduces the value of 
$\langle 0|\rho |1\rangle$ \cite{book,zurek,joos}, since
the bath entangles with the quantum dot, and destroys 
the quantum correlations between the dot and the
conduction electrons. 
When ${\cal V}$ is strong
enough to make $\langle 0|\rho |1\rangle$ negligible, we have 
$\rho_f= a_1 |0\rangle\langle 0|+ a_2 |1\rangle\langle 1|)$ and,
as shown above, the Kondo effect is destroyed.

Destruction of the Kondo effect by a decohering action of an
external microwave field has been considered in Ref.\
\cite{microwave}. However, the external radiation works similar
to increasing temperature, smearing the Kondo effect, and
represents an effect of dissipation. In contrast to dissipation,
the decoherence caused by the measurement leads to the pure
decrease of the amplitude of the resonance (in $1/Z=1-\langle
n_f\rangle$ times), without any smearing. The difference between
decoherence and dissipation is one of the most important
points in the modern theory of decoherence \cite{zurek}.

Here, we consider decoherence by a spin bath \cite{hyper}, 
which allows a clear demonstration of the
dissipationless suppression of the Kondo effect. 
Such a bath can be implemented in experiments by
doping GaAs with manganese (so that the magnetic moments of Mn
ions form the bath) or chromium impurities. The interaction between
the spins of impurities is weak but not negligible:
it determines the chaotic (or close to chaotic)
dynamics of the bath, and our results show
that this is a qualitatively important detail. 
Rigorous treatment of this problem is very difficult, but 
qualitative features of decoherence of a Kondo system can be 
studied by representing the spin state of the subsystem of 
conduction electrons
by a single collective spin 1/2. I.e., instead of the
Hamiltonian (\ref{kasham}), we consider a qualitatively
similar Hamiltonian
\begin{equation}
H=J{\bf S}_1{\bf S}_2 + {\bf S}_1\sum_{j=1}^{N} A_j{\bf I}_j + H_B
\label{hyperfine}
\end{equation}
which describes essentially the same physics as (\ref{kasham}).
Here, ${\bf S}_1$ is the spin of the quantum dot 
($|{\bf S}_1|=1/2$), ${\bf S}_2$ represents the collective spin 
of the conduction electrons ($|{\bf S}_2|=1/2$), 
${\bf I}_j$ are the environmental spins ($|{\bf I}_j|=1/2$),
and $A_j$ are the coupling constants
of the spin of the dot ${\bf S}_1$ with the bath spins. 
The Hamiltonian $H_B$ describes the chaotic dynamics of the bath.

Both Hamiltonians (\ref{kasham}) and (\ref{hyperfine})
predict a singlet ground state in the absence of the bath;
the energy scales of the Hamiltonians match, since the parameter 
$J>0$
is the {\it renormalized\/} (effective) coupling of the spin of 
the dot with the conduction electon subsystem,
i.e., $J=E^{\ast}\sim T_K$ (so $J\ll J_{sd}$).
The entanglement between 
${\bf S}_1$ and ${\bf S}_2$ can be described using the
reduced density matrix $\rho = \mathop{\rm Tr}_{\{{\bf I}_j\}} W$,
where $W$ is the density matrix of
of the whole system (the two central spins ${\bf S}_{1,2}$
plus the all the bath spins), and the trace is taken over
the bath spins \{${\bf I}_j$\}. The entangled singlet state of the
quantum dot and the conduction electrons (where the
Kondo effect is maximal) corresponds to 
the non-diagonal element 
$\rho_{12}=\langle\uparrow\downarrow|\rho|%
\downarrow\uparrow\rangle=-1/2$. 
The decay of entanglement between 
${\bf S}_1$ and ${\bf S}_2$ is characterized by a decrease
of the absolute value of $\rho_{12}$. When $\rho_{12}$
vanishes, the Kondo effect disappears.
%is essentially weakened.

\begin{figure}
\includegraphics[width=9cm]{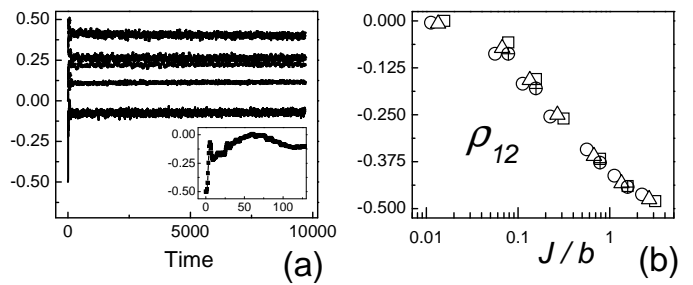}
\caption{Decoherence dynamics of the Kondo system by a
spin bath. (a) --- temporal evolution of the different
elements of the density matrix $\rho$: diagonal elements
corresponding to the states $|\uparrow\uparrow\rangle$,
$|\uparrow\downarrow\rangle$, $|\downarrow\uparrow\rangle$,
and $|\downarrow\downarrow\rangle$
(four upper curves),
and the non-diagonal element $\rho_{12}$ (lowest curve).
The inset shows initial decrease in $\rho_{12}$.
(b): dependence of the final value of $\rho_{12}$ vs.
the ratio $J/b$; the results for different 
$N$ and $A_j$ are indicated by different symbols.
Different results are close to the same universal curve.}
\label{figure}
\end{figure}

We study this process by direct numerical solution \cite{hans}
of the compound ``system-plus-bath'' time-dependent
Schr\"odinger equation with the Hamiltonian
(\ref{hyperfine}). 
Initially, the system and the bath are in an uncorrelated
product state. The initial state of the Kondo system is
the singlet. The initial state
of the bath is a random superposition
of all basis states,
which corresponds to low-temperature experiments
when the temperature 
$J\gg T\gg A_j$. The chaotic dynamics of the bath has been
simulated by using the form of $H_B$ suggested in
\cite{shepel}; the level statistics has been checked to
agree with the Wigner-Dyson distribution.
The number of the bath
spins has been varied from $N=12$ to $N=6$, and different
sets of $A_j$ have been used.
The results of a sample run are shown in Fig.\ \ref{figure}(a).
Decoherence dynamics of the Kondo system and the decay
of the element $\rho_{12}$ are clearly seen.

The simulations illustrate the dynamics of 
the measurement process, which starts from
the singlet state of the Kondo system. For the quantum
dot with magnetic impurities, this dynamics is 
not relevant (since the initial product state
is not realistic),
but the final quasi-equilibrium state is absolutely
meaningful, giving the correct value of $\rho_{12}$.
This value, as well as general
features of the system's evolution, is stable with
respect to considerable changes in the parameters of $H_B$,
number of spins, values of $A_k$, or variation of the
initial conditions. Thus, the final state of the
system represents the ``pointer state'' \cite{book,zurek}
which is robust with respect to decoherence.

The equilibrium value of $\rho_{12}$ is determined by
a competition between the exchange constant
$J$ and the strength of the system-bath interaction.
Analysis similar to \cite{hyper}
suggests that the relevant quantity characterizing 
the system-bath coupling is the mean-square exchange
$b=\sqrt{\sum_j A_j^2}$, so the final value of $\rho_{12}$
is determined by the ratio $J/b$. Our results show
that this is correct (see Fig.\ \ref{figure}(b)):
the results obtained with different number of the bath
spins $N$, different sets of $A_j$, and different 
values of $J$, fall close to a universal curve $\rho_{12}(J/b)$.
The scatter is moderate, stemming from the finite value
of $N$ and fluctuations present in the final state
(Fig.\ \ref{figure}(a)).

The suppression of the Kondo
effect as a function of $J/b$ is gradual, and the
center of the transition corresponds to $J/b\approx 0.3$.
For typical quantum dots \cite{qdexp}, $J\sim T_K\approx 0.4$ K,
and $A_k\sim I/n$ where $n\sim 10^7$ is the
total number of the lattice sites inside the dot
and $I\sim 1$ eV is the exchange of the impurity spin
with the electron of the dot (the factor $1/n$ originates
from the normalization of the wave function of the 
electron in the quantum dot). Thus, $b\sim \sqrt{n_i} I/n$
where $n_i$ is the number of impurities in the dot.
At large concentrations $x=n_i/n\sim 0.01$, 
Mn impurities in GaAs order ferromagnetically \cite{ohno},
and our model becomes
invalid (impurities no longer form a spin bath), but
for other ions, such as Fe \cite{qdFe}, the critical
concentration is larger, and $x\sim 0.01$ is acceptable.
Therefore, in realistic systems
$b\sim 0.1$--0.3 K, and the ratio $T_K/b \sim 0.3$
is easily achievable, so that an experimental check
of our predictions is possible. The experiment
is rather straightforward: several Fe- or Mn-doped samples 
with different impurities concentration
should be prepared, 
and the Kondo-anomaly should be measured, similar to
\cite{qdexp}. The Kondo effect can be suppressed further by
reducing the size of the dot (since
the ratio $J/b$ is proportional to $n$, i.e. to the volume
of the dot).

Note that doping of the dot with non-magnetic impurities
having internal degrees of freedom, e.g.\ with
Ce atoms, will suppress the Kondo effect in exactly the same 
manner.
Another possibility is to use a double quantum dot
system \cite{double}, where the Kondo effect changes due to
the presence of the ``orbital''
(right or left dot) degree of freedom, in addition to the
spin of the dots. The measurement of the electron
presence in a given dot \cite{zeno} will bring the Kondo
effect to the single dot regime.

Summarizing, we have shown that the quantum measurement
of the spin in a Kondo system suppresses the Abrikosov-Suhl
resonance and destroys the Kondo effect. This suppression
is caused by the decohering influence of the measuring apparatus,
and does not involve dissipation, i.e.\ it is 
qualitatively different from the dissipative suppression
of the Kondo effect \cite{microwave}. The effect predicted here
can be studied in realistic experiments on quantum dots doped with
magnetic (Mn, Cr) or non-magnetic (Ce) impurities, where
the bath of impurities decoheres the Kondo system in the same
way as a measuring device. The estimates show that such
an experiment is already achievable with today's experimental 
techniques.

This work was partially carried out at the Ames Laboratory, which
is operated for the U.\ S.\ Department of Energy by Iowa State
University under Contract No.\ W-7405-82 and was supported by the
Director of the Office of Science, Office of Basic Energy
Research of the U.\ S.\ Department of Energy. Support from the
Dutch ``Stichting Nationale Computer Faciliteiten (NCF)''
is gratefully acknowledged. This work was partially supported by
Russian Science Support Foundation.

\end{document}